\documentclass[authoryear,review]{elsarticle1}
\def\astrobj#1{#1}

\begin{document}

\begin{frontmatter}

\title{Optical Spectroscopy of Four Young Radio Sources}

\author[1,2,4]{Xu-Liang Fan}
\author[1,2]{Jin-Ming Bai}
\author[3]{Chen Hu}
\author[1,2]{Jian-Guo Wang}

\address[1]{Yunnan Observatories, Chinese Academy of Sciences, Kunming 650011, China;  fxl1987@ynao.ac.cn, baijinming@ynao.ac.cn}
\address[2]{Key Laboratory for the Structure and Evolution of Celestial Objects, Chinese Academy of Sciences, Kunming 650011, China}
\address[3]{Key Laboratory for Particle Astrophysics, Institute of High Energy Physics, Chinese Academy of Sciences, 19B Yuquan Road, Beijing 100049, China}
\address[4]{University of Chinese Academy of Sciences, Beijing 100049, China}

%     Abstract is required.

\begin{abstract}
We report the optical spectroscopy of four young radio sources which are observed with the Lijiang 2.4m telescope. The Eddington ratios of these sources are similar with those of narrow-line Seyfert 1 galaxies (NLS1s). Their Fe {\sc ii} emission is strong while [O {\sc iii}] strength is weak. These results confirm the NLS1 features of young radio sources, except that the width of broad H$\beta$ of young radio sources is larger than that of NLS1s. We thus suggest that the young radio sources are the high black hole mass counterparts of steep-spectrum radio-loud NLS1s. In addition, the broad H$\beta$ component of \astrobj{4C 12.50} is the blue wing of the narrow component, but not from the broad line region.
\end{abstract}

\begin{keyword}
galaxies: jets --- quasars: emission lines --- quasars: individual (\astrobj{GB6 J0140+4024}, \astrobj{TXS 0942+355}, \astrobj{IRAS 11119+3257}, \astrobj{4C 12.50})
\end{keyword}

\end{frontmatter}

\section{Introduction}           %% first-level sections will be auto-capitalized
\label{sect:intro}
The young radio sources, including compact steep spectrum (CSS) and gigahertz peaked spectrum (GPS) radio sources, are believed to represent the earliest stages in the evolution of the powerful radio galaxies \citep{1998PASP..110..493O}. Some of them are found to be associated with galaxies mergers \citep{1986Natur.321..750G} or ultraluminous infrared galaxies (ULIRGs) \citep{2011MNRAS.410.1527H, 2012MNRAS.422.1453N}. Recent observations also manifest that some radio-loud narrow-line Seyfert 1 galaxies/quasars (NLS1s) have the radio properties consistent with CSS radio sources \citep{2010AJ....139.2612G, 2014MNRAS.441..172C, 2015arXiv151005584L}. NLS1s are classified based on their narrow Balmer lines (with the full width at half maximum, FWHM $< 2000~km~s^{-1}$), small ratio between [O {\sc iii}]$\lambda$5007 and H$\beta$ ([O {\sc iii}]$\lambda$5007/H$\beta$ $<$ 3) and strong emission of Fe {\sc ii} complexes (Fe {\sc ii} $\lambda$4570/H$\beta$ $>$ 0.5, \citealt{1985ApJ...297..166O, 2001A&A...372..730V}). These features are explained as a result of relatively small mass of the central black hole with high accretion rate (\citealt{2002ApJ...565...78B, 2012AJ....143...83X} , but also see \citealt{2008MNRAS.386L..15D}).  Thus NLS1s are suggested to be during the early stage of the accretion activities.

The accretion rates of young radio sources are found to be relatively high, with the average value similar with NLS1s \citep{2009MNRAS.398.1905W, 2016ApJ...818..185F}. Thus the young radio sources can also stand during the early stage of accretion activities. Moreover, there is another similar feature between the young radio sources and NLS1s. That is the blue wing of narrow [O {\sc iii}] \citep{2005MNRAS.364..187B, 2008MNRAS.387..639H, 2009MNRAS.398.1905W, 2009MNRAS.400..589H}. This feature is always explained as the outflow originated from the disk wind or galactic wind. Some results indicate that the strength of the blueshift is related to the Eddington ratio \citep{2008ApJ...680..926K}, while other explanations refer to the jet - ISM interaction \citep{2008MNRAS.387..639H}.

Although the common radio properties between NLS1s and young radio sources have been discussed frequently in the literature, the common optical properties between them are less discussed. In this paper, we obtain the optical spectra of four young radio sources, and explore their emission lines properties. Throughout this paper, the luminosity is calculated using a $\Lambda$CDM cosmology model with h=0.71, $\Omega_{m}$=0.27, $\Omega_{\Lambda}$=0.73.

\section{Observations and Data Reduction}
\label{sect:Obs}

The spectra were obtained between 2014 and 2015, with the Yunnan Faint Object Spectrograph and Camera (YFOSC) on the Lijiang 2.4m telescope in Yunnan Observatories. The YFOSC is equipped with a back-illuminated, blue sensitive CCD with 2048 $\times$ 4608 pixels, which works in both imaging and spectroscopy modes. Each pixel of the YFOSC corresponds to the sky angle $0.283^{''}$. The spectra were taken with two gratings, G8 and G3. The dispersions of G8 and G3 are 1.5 $\AA~pixel^{-1}$ and 2.9 $\AA~pixel^{-1}$, respectively. The wavelength coverage is about 4970 - 9830 \AA~and 3200 - 9200 \AA~for G8 and G3, respectively. The details of observations are listed in Table ~\ref{tab1}.
\begin{table}[H]
\caption[]{The Observation Log \label{tab1}}
\vspace{-1mm}\footnotesize
 \begin{center}\doublerulesep 0.1pt \tabcolsep 0.5pt
 \begin{tabular}{ccccccc}
  \hline\noalign{\smallskip}
 Source Name & Date & Exp. (s) & slit ($''$) & Grim & $\lambda$ Range (\AA) & Res. (\AA)\\
  \hline\noalign{\smallskip}
 \astrobj{GB6 J0140+4024} & 2014/11/25 & 2400 & 1.8 & G3 & 4264 - 8603 & 16 \\
 \astrobj{TXS 0942+355} &  2015/03/13 & 3000 & 2.5 & G3 & 3892 - 9110 & 20 \\
 \astrobj{IRAS 11119+3257} & 2015/03/16 & 3000 & 2.5 & G3 & 4156 - 9109 & 20 \\
 \astrobj{4C 12.50} & 2014/05/30 & 2400 & 1.8 & G8 & 5182 - 9518 & 8 \\
  \noalign{\smallskip}\hline
\end{tabular}
\end{center}
\end{table}

\begin{figure}
   \centering
  \includegraphics[width=8.5cm, angle=0]{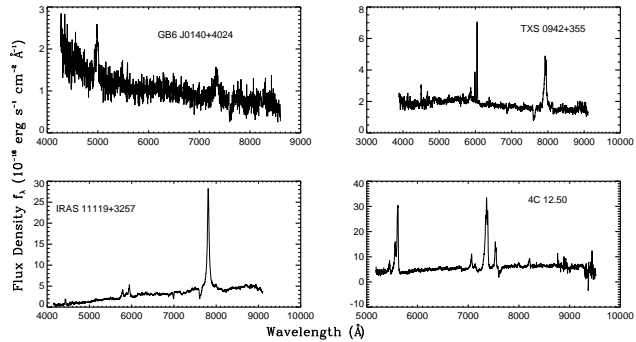}
   \caption{The spectra of young radio sources.}
   \label{fig1}
\end{figure}

The spectra are reduced with the standard IRAF routines, including the bias substraction, flat field corrections and the removal of cosmic rays. Wavelength calibration is performed with the neon and helium lamps. The spectra are flux calibrated with the spectrophotometric flux of standard stars observed at a similar air mass on the same night.  The spectral resolution is estimated using the FWHM of the sky emission lines (Table ~\ref{tab1}). The four reduced 1-d spectra are plotted in Figure ~\ref{fig1}.

\section{Results on Individual Objects}
The galactic extinction is firstly corrected using the dust map from \citet{1998ApJ...500..525S} and the extinction curve from \citet{1989ApJ...345..245C} with $R_V$ = 3.1. Then we fit the four spectra in the rest frame. The pseudo-continuum are decomposed with a power-law AGN continuum, a host galaxy component (for the optical band in the rest frame), and a Fe {\sc ii} template. The fitting algorithm is followed Hu et al. \citep{2015ApJ...804..138H} (also see \citealt{2012ApJ...760..126H, 2009ApJ...707.1334W, 2011ApJS..194...45S}). The emission lines are fitted with Gaussian profiles or Gauss - Hermite function. Throughout this paper, we just focus on the H$\beta$ and Mg {\sc ii} region. The H$\beta$ and Mg {\sc ii} are treated as a broad component plus a narrow one. If one single Gaussian can not fit the broad component well, we model it with a Gauss - Hermite function (for \astrobj{IRAS 11119+3257}). Each of [O {\sc iii}] doublet is fitted with one Gaussian. For \astrobj{4C 12.50}, two Gaussian profiles are performed to model the unusual [O {\sc iii}]. Then the fitted FWHM is corrected for the instrumental broaden listed in Table \ref{tab1}. The strength of optical Fe {\sc ii} is represented by Fe {\sc ii} $\lambda$4570 (integrated between $\lambda$4434 and $\lambda$4684). The ultraviolet (UV) Fe {\sc ii} is integrated in the range 2200 - 3090 \AA. The details of each sources are discussed below.

\begin{figure}
   \centering
  \includegraphics[width=8.5cm, height=6cm, angle=0]{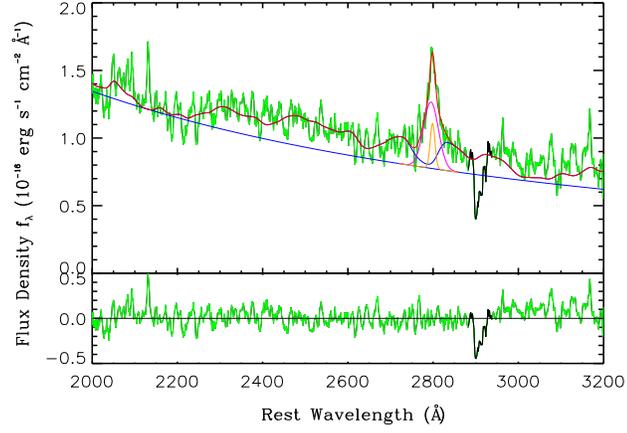}
   \caption{The spectra of \astrobj{GB6 J0140+4024}. The top panel shows the rest frame spectrum which corrected Galactic extinction (green), along with the fitting results (red). The black points present the data out of fitting. The fitting model includes a pseudo-continuum (blue, composed of the AGN power-law continuum and UV Fe {\sc ii} emission) and Mg {\sc ii} emission lines (broad component in magenta, and narrow component in orange). The bottom panel shows the residuals.}
   \label{fig2}
\end{figure}

\textbf{\astrobj{GB6 J0140+4024}}. This source encounter a low signal/noise ratio (S/N). However, the Mg {\sc ii} and C {\sc iii} lines are prominent (upper left panel in Figure \ref{fig1}). The spectrum in rest frame which is corrected the Galactic extinction around Mg {\sc ii} is plotted in Figure \ref{fig2}, along with the modeled components. The flux and FWHM of the broad Mg {\sc ii} are $21.59 \pm 3.10 \times 10^{-16}~erg~s^{-1}$ and $4389 \pm 714~km~s^{-1}$, respectively. The flux of Fe {\sc ii} is $100.86 \pm 24.17 \times 10^{-16}~erg~s^{-1}$.

\begin{figure}
   \centering
  \includegraphics[width=6cm, height=8.5cm, angle=-90]{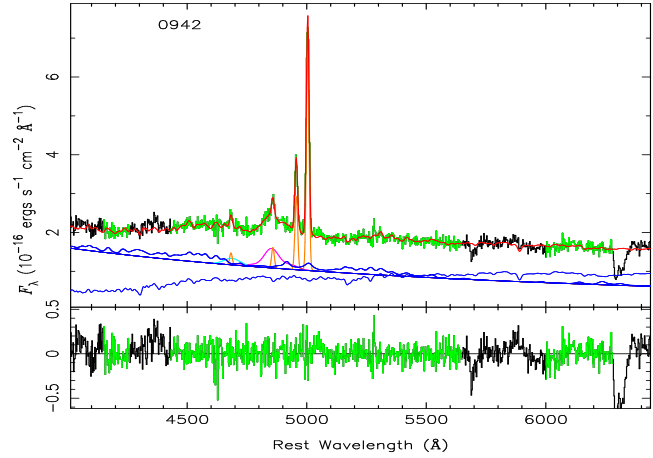}
   \caption{The spectra of \astrobj{TXS 0942+355}. The manners are same as those in Figure 2, but the pseudo-continuum is also composed of a host galaxy component. The broad H$\beta$ is in magenta, broad He {\sc ii} $\lambda$ 4686 is in cyan, and narrow emission lines are in orange.}
   \label{fig3}
\end{figure}

\textbf{\astrobj{TXS 0942+355}}. \citet{2010MNRAS.408.2261K} labelled this sources as a low-luminosity CSS source. \citet{2010MNRAS.408.2279K} noted that the optical image showed weak extended emission. The optical spectrum manifests a large contribution of host galaxy component (Figure \ref{fig3}). We obtain the flux and FWHM of the broad H$\beta$ which are $50.52 \pm 3.52 \times 10^{-16}~erg~s^{-1}$ and $4726 \pm 605~km~s^{-1}$, respectively. The [O {\sc iii}]$\lambda$5007/H$\beta$ and Fe {\sc ii}/H$\beta$ (the so called R4570 parameter) are 1.59 and 0.8, respectively.

\begin{figure}
   \centering
  \includegraphics[width=6cm, height=8.5cm,angle=-90]{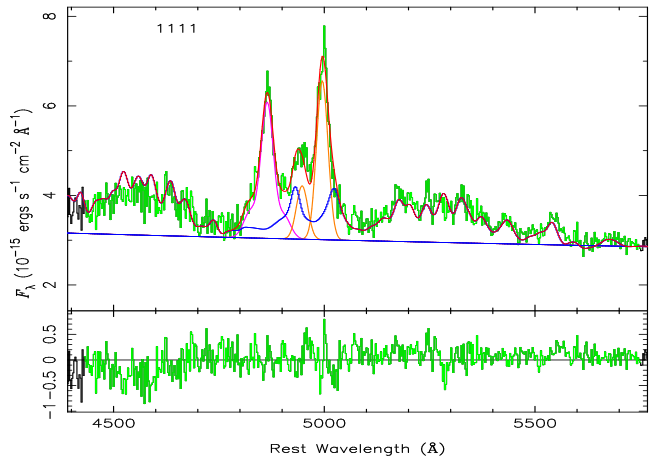}
  % \begin{minipage}[]{85mm}
   \caption{The spectra of \astrobj{IRAS 11119+3257}. The manners are same as those in Figure 3.}
%\end{minipage}
   \label{fig4}
\end{figure}
\textbf{\astrobj{IRAS 11119+3257}}.  This is a famous ULIRG with high velocity outflows observed at optical and X-ray band \citep{2003MNRAS.340..289L, 2015Natur.519..436T}. \citet{2006AJ....132..531K} labelled this sources as a radio-loud NLS1 and presented that its radio feature was compact and steep spectrum. The optical spectrum of \astrobj{IRAS 11119+3257} show large intrinsic extinction \citep{2002AJ....124...18Z}. Thus a intrinsic extinction $A_V = 2.5$ mag is corrected to make the Balmer decrement close to 3.1 \citep{2008MNRAS.383..581D}.  The peak of [O {\sc iii}] has a blueshift about 700 km s$^{-1}$ relative to the peak of H$\beta$. The H$\beta$ can be well fitted with a Gauss - Hermite function. Thus we treat total H$\beta$ as a broad component, while the narrow component of H$\beta$ is too weak to model \footnote{Two Gaussian profiles are also used to model H$\beta$. The central wavelengths of the two Gaussians also have systematic offsets with that of [O {\sc iii}].} (Figure \ref{fig4}). The FWHM of broad H$\beta$ is $2070 \pm 110~km~s^{-1}$, which is slightly larger than  $1980~km~s^{-1}$ listed in \citet{2002AJ....124...18Z}. The flux of broad H$\beta$ is $1520.5 \pm 42.8 \times 10^{-16}~erg~s^{-1}$. The [O {\sc iii}]$\lambda$5007/H$\beta$ and Fe {\sc ii}/H$\beta$ are 0.80 and 1.58, respectively.

\begin{figure}
   \centering
  \includegraphics[width=6cm, height=8.5cm, angle=-90]{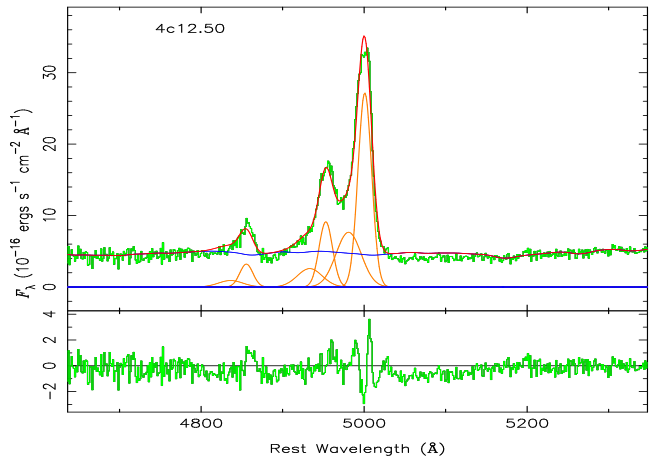}
   \caption{The spectra of \astrobj{4C 12.50}. The manners are same as those in Figure 3.}
   \label{fig5}
\end{figure}

\textbf{\astrobj{4C 12.50}}. \astrobj{4C 12.50} (PKS 1345+12) is hosted in a major merger system which is still ongoing \citep{1986Natur.321..750G}. The GPS nuclei was classified as a narrow line radio galaxy (NLRG) by \citet{1977ApJ...215..446G}. However, it has very broad [O {\sc iii}] lines. The broad [O {\sc iii}] can be fitted by two Gaussian profiles (Figure \ref{fig5}). Recent analysis of its SDSS spectrum showed a broad H$\beta$ component \citep{2012ApJ...757..140S}. Our results show that the broad H$\beta$ has the same blueshift (about 1549 km s$^{-1}$) with the blueshifted [O {\sc iii}] doublet. And its FWHM (2130 km s$^{-1}$) is also consistent with the blue wing of [O {\sc iii}]. Thus the detected broad H$\beta$ corresponds to the blue shifted narrow line component, but not originates from broad line region (BLR).

\section{Discussion}
\label{sect:discussion}
We estimate the black hole mass of three type 1 sources. The calculations for H$\beta$ follow the relation of single epoch reverberation mapping in \citet{2006ApJ...641..689V},
\begin{equation}
  log M_{BH} = 2~log(\frac{FWHM (H\beta)}{1000~km~s^{-1}}) + 0.5~log(\frac{\lambda L_{\lambda}(5100)}{10^{44}~erg~s^{-1}}) + 6.91 .
\end{equation}
For Mg {\sc ii}, we use the relation in \citet{2009ApJ...707.1334W},
\begin{equation}
  log M_{BH} = 1.51~log (\frac{FWHM (Mg~{\sc II})}{1000~km~s^{-1}}) + 0.5~log(\frac{\lambda L_{\lambda}(3000)}{10^{44}~erg~s^{-1}}) + 7.13 .
\end{equation}
Then we estimate the Eddington ratio $L_{bol}/L_{Edd} = 9 L_{5100}/(1.3\times 10^{38} M_{BH})$ \citep{2000ApJ...533..631K}. The results are listed in Table \ref{tab2}.

\begin{table*}
\caption[]{The black hole mass and Eddington ratio. The $\lambda$5100 luminosity of \astrobj{GB6 J0140+4024} is calculated from $\lambda$3000 with the spectral index -1.65. The black hole mass and Eddington ratio of \astrobj{4C 12.50} are taken from \citet{2006ApJ...638..745D} and \citet{2011MNRAS.410.1527H}. \label{tab2}}
\vspace{-1mm}\footnotesize
 \begin{center}\doublerulesep 0.1pt \tabcolsep 0.5pt
 \begin{tabular}{cccccc}
  \hline\noalign{\smallskip}
 Source Name & redshift & $L_{5100}$ ($erg~s^{-1}$) & Log$M_{BH}$ & $L_{bol}/L_{Edd}$ \\
  \hline\noalign{\smallskip}
 \astrobj{GB6 J0140+4024} & 1.62 & $2.58\times 10^{45}$ & 8.88 & 0.23 \\
 \astrobj{TXS 0942+355} & 0.208 & $6.28\times 10^{43}$ & 7.66 & 0.10 \\
 \astrobj{IRAS 11119+3257} & 0.189 & $1.51\times 10^{45}$ & 7.63 & 2.45 \\
 \astrobj{4C 12.50} & 0.122 & --- & 7.82 & 0.27 \\
  \noalign{\smallskip}\hline
\end{tabular}
 \end{center}
\end{table*}

The estimated average values of Eddiongton ratios for NLS1s are from 0.15 \citep{2008MNRAS.390..752B} to 0.79 \citep{2012AJ....143...83X}, while the broad line Seyfert 1 galaxies (BLS1) have the average value about 0.16 \citep{2012AJ....143...83X}. The Eddington ratios of four young radio sources are generally larger than BLS1 and distributes in the range of NLS1s.

The three type 1 sources all show strong Fe {\sc ii} emissions, and [O {\sc iii}] emission is weak for \astrobj{TXS 0942+355} and \astrobj{IRAS 11119+3257}.  These features confirm that the emission lines properties of young radio sources are similar with NLS1s, except that the line width of young radio sources is broader than that of NLS1s. The estimated black hole mass is also larger than the average value of NLS1s \citep{2012AJ....143...83X}. Meanwhile, the radio powers of the compact steep-spectrum NLS1s are at the low end of that of young radio sources \citep{2010AJ....139.2612G, 2014MNRAS.441..172C, 2015arXiv151005584L}, but locate in the range of the low-luminosity compact radio sources ($P_{1.4GHz} < 10^{26}~W~Hz^{-1}$, \citealt{2010MNRAS.408.2261K}). Therefore, we suggest the young radio sources are the high mass counterparts of the steep-spectrum radio-loud NLS1s.

The blue wing of [O {\sc iii}]  is a prominent feature in NLS1s \citep{2005MNRAS.364..187B, 2008ApJ...680..926K}. Among the four sources, the blue wings of [O {\sc iii}] are presented  in \astrobj{IRAS 11119+3257} and \astrobj{4C 12.50}. Both sources are ULIRGs, and with relatively high Eddiongton ratio, which links the outflow mechanism to high star formation rate or high accretion rate.

\section{Acknowledgments}
We acknowledge the support of the staff of the Lijiang 2.4m telescope. Funding for the telescope has been provided by CAS and the People's Government of Yunnan Province. This research is supported by the Strategic Priority Research Program of the Chinese Academy of Sciences - The Emergence of Cosmological Structures (grant No. XDB09000000), the Key Research Program of the Chinese Academy of Sciences (grant No. KJZD-EW-M06), and the NSFC through NSFC-11133006 and 11361140347.

\end{document}